\documentstyle[aps,prl,epsf]{revtex}
\newcommand{\beq}{\begin{equation}}
\newcommand{\eeq}{\end{equation}}

\twocolumn
\begin{document}
\draft
\twocolumn[\hsize\textwidth\columnwidth\hsize\csname @twocolumnfalse\endcsname

\title{ Lack of Self Averaging and Finite Size Scaling in Critical Disordered 
Systems}

\author{Shai Wiseman and Eytan Domany }
\address{Department of Physics of Complex Systems, \\
         The Weizmann Institute of Science,
         Rehovot 76100, Israel
}

\date{Submitted to Phys. Rev. Lett., Feb. 9, 1998}
\maketitle
\begin{abstract}
We simulated site dilute Ising models in $d=3$ dimensions for
several lattice sizes $L$. 
For each $L$  singular thermodynamic quantities $X$
were measured at criticality and their distributions $P(X)$ were determined, 
for  ensembles of several thousand 
random samples.
For $L \rightarrow \infty$ the width of $P(X)$
tends to a universal constant, i.e. 
there is no self averaging. 
The width of the distribution 
of the sample dependent  pseudocritical temperatures $T_c(i,L)$  
scales as $\delta T_c(L)  
\sim L^{-\frac{1}{\nu}}$ and {\em not} as 
$\sim L^{-\frac{d}{2}}$. Finite size scaling holds;  
the sample dependence of $X_i(T_c)$ enters predominantly 
through 
$T_c(i,L)$.

\end{abstract}
]
Phase transitions in systems with quenched disorder are of considerable
theoretical and experimental interest and have been the 
subject of intensive investigations. 
Nonetheless, there are only few general exact results on such systems
 and some of
these raised questions which were left unanswered for more than a decade.
For example, the Harris criterion\cite{Harris 1974} states that the condition for the stability
of a pure fixed point against disorder is $1/\nu_{\text{pure}} \leq d/2$.
This criterion was derived assuming that fluctuations of the local
critical temperature $T_{c,l}$ of a region of size $\xi_l^{d}$ scaled as 
$\delta T_{c}\sim \xi^{-d/2}$.
Another result by Chayes et al\cite{Chayes 1986} stated that $1/\nu \leq d/2$ for any disordered
 system. This result is identical with the Harris criterion for disordered
 systems governed by a pure fixed point. For systems governed by a random 
fixed point it raised the question whether the relation 
 $\delta T_{c}\sim \xi^{-d/2}$  holds in this case as well?


Another important and relevant issue concerns the meaning of
measurements done on disordered systems at or near their transition points.
All measurements (experimental and numerical) are obtained for finite
systems; moreover, each such system constitutes a particular realization
of the quenched randomness. Hence it is natural to ask 
to what extent are  results obtained for a single system representative of 
the general class to which it belongs? The answer to this question
hinges on the important issue of self-averaging.
If a quantity is not self averaging, increasing its size $L$ does not
improve the statistics of its measurement (sample to sample fluctuations
remain large). 
Whereas it has been known that for (spin and
regular) glasses\cite{Rev:Mod} there is no  self-averaging in the ordered phase, the 
discovery\cite{wd95} that there is no self averaging for random {\it ferromagnets}
at their critical point came as somewhat of a surprise. 
The numerical work on which this claim was based has been backed up with a
finite-size scaling ansatz\cite{wd95} which seemed to fit the data quite well.
Subsequently the issue has been investigated by Aharony and Harris\cite{ah96}
 (AH), who
provided a theoretical understanding of the absence of self averaging 
in critical random ferromagnets. 
The central point of the AH argument was that a random
fixed point is characterized by a {\it distribution of non-zero width} 
of some measurable
quantities (such as pseudo-critical temperatures, susceptibilities etc)
and hence 
if an ensemble of systems flows to a such a fixed point, its corresponding 
properties
must also be distributed in a similar way.
The general argument was supported by a Renormalization Group 
calculation in $d=4 - \epsilon$ dimensions.
The AH work led to several predictions, some of  which were in clear
disagreement with the scaling theory\cite{wd95}. Since neither the
general arguments nor results obtained by
$\epsilon$-expansion can be viewed as definitive, independent 
confirmation is desired. Furthermore, recently claims were made\cite{hun97} 
to the effect
that the manner in which the sample to sample fluctuations of the 
pseudoritical
temperature scales with size governs one's ability to observe the 
"true" critical
exponents of random systems, quantum and classical. 

In order to investigate these fluctuations, 
the extent to which our finite size scaling ansatz holds and to 
confirm or disprove the AH results we carried out extensive 
simulations of the $3-d$ Ising model with site dilution. 
Since for the pure model the specific heat exponent 
$\alpha_p = 0.11 >0$\cite{Fisher74},
randomness is relevant and the critical behavior is governed by
a random fixed point. Various aspects of the critical properties of
this model have been measured very carefully\cite{heuer}. 
In particular, $\alpha$, 
the specific heat exponent of the random model is negative\cite{heuer},
consistently with \onlinecite{Chayes 1986}.

To state 
the main results of our study we first recapitulate a few definitions.
We consider an ensemble of systems of linear size $L$, denoting 
a particular realization of the randomness by $i$. For each such system
we measure, at temperature $T$, various thermodynamic densities 
$X_i(T,L)$ (such as the 
susceptibility per site $X=\chi$, magnetization $m$, etc). 
Denote by $T_c(i,L)$ 
the pseudocritical temperatures of each of these samples (obtained, for
example, by locating the maximum of $\chi$, see Fig 1). 
The sample-averaged pseudocritical temperature,
$T_c(L)=[T_c(i,L)]$ approaches, for $L\rightarrow \infty$,  
the asymptotic limit 
$T_c(L) \rightarrow T_c$. 
Our main results
can be summarized as follows:
\begin{enumerate}
\item \label{item lack}
{\it Lack of self-averaging:} 
The values of $\chi$,
measured  at $T_c$ for many
samples of size $L$,
are distributed with mean 
$[\chi (L)]$ and variance
$V_{\chi}(L)$.  The normalized square width 
\[ R_{\chi}= V_{\chi}/[\chi]^2 \rightarrow C
 \;\;\mbox{as}\;\;L\rightarrow \infty\]
{\it i.e. goes to a constant, for large $L$}, as predicted by AH, in 
contradiction to
our prediction $R \sim L^{\alpha/\nu}$.
The 
distributions of various thermodynamic quantities do not become sharp in
the thermodynamic limit and the system is not self-averaging.
\item
{\it Universality:} AH predicted that the constant $C$ is universal. To test
this, we studied different dilutions and two random
ensembles; one
with a fixed {\it concentration} of magnetic sites (grand canonical) and the
other with a fixed {\it number} of spins, placed at random on the lattice 
(canonical). Grand canonical ensembles with different concentrations indeed 
have the same asymptotic width, but two different ensembles seem to have 
different values\cite{ahw98,wd98} of $R$
(even when the concentration
in one equals the fraction of spins of the other).

\item
The pseudocritical temperatures $T_c(i,L)$ are distributed with a width 
$ \delta T_c(L) $ which
scales as
\beq
\delta T_c(L) \sim L^{-\frac{1}{\nu}}
\label{eq:AHwidth}
\eeq
This behavior was sugested by AH, reasoning that when combined with the
finite size scaling theory of \onlinecite{wd95} it gave rise to the lack of
 self averaging discussed in item \ref{item lack}. However this reasoning
 depended on the correctness of the finite size scaling theory.
Here we have directly shown the validity of (\ref{eq:AHwidth}) independently
 of any additional assumptions. We rule out the possibility that 
$\delta T_c(L) \sim L^{-d/2}$ as was
assumed by us\cite{wd95} and listed by others\cite{hun97} as "the most likely
scenario".  The latter behavior occurs only in systems governed by a pure
fixed point where the Harris criterion is valid. Note that the critical 
concentration in percolation\cite{Stauffer 1992} behaves as in 
(\ref{eq:AHwidth}) (see also a similar phenomenon in the random field
Ising model\cite{Sourlas 1997}).

\item    \label{item: fss}
The finite size scaling form assumed in\cite{wd95},
\beq
X_i(T,L) \approx L^\rho Q_i({\dot t_i}L^\frac{1}{\nu})
\label{eq:FSS}
\eeq
where ${\dot t}_i$ is a sample-dependent reduced temperature,
\beq
{\dot t_i} = [T - T_c(i,L)]/T_c
\label{eq:tdot}
\eeq
is indeed correct. Particularly, the sample dependence of $X_i$ is 
predominantly due to the sample dependence of $\dot{t}_i$ and not due to the 
sample dependence of the function $Q_i$.

\item
However, the scaling function $Q_i(x)$ {\it does} depend on the realization $i$.

\item
It may be computationally advantageous to measure various quantities,
for instance, to find critical exponents, at
the sample-dependent pseudocritical point $T_c(i,L)$ (where $\dot{t}_i=0$),
 rather than at $T_c$.
 This is so since, as a consequence of item \ref{item: fss}, 
the variance at $T_c(i,L)$  is much smaller. 
\end{enumerate}
We present now numerical evidence for each of the statements made above.

{\it The site-dilute Ising model:} Each site of a cubic lattice is either 
occupied by an Ising spin or empty. Here we report results obtained for a
grand-canonical ensemble of
samples in which the occupation of each site was determined independently
with probability $p=0.8$. At this probability Heuer found the fastest 
crossover to the asymptotic (random) critical behavior\cite{heuer}. 
We used  the Wolff\cite{Wol:1C} single cluster
algorithm\cite{Wang90a} with skewed periodic boundary 
conditions\cite{BH:book} on lattices of sizes $L=4,8,16,32,64$. For each 
of these sizes
we simulated, respectively,
 \[ n_L=10000, 4000,32000,4000,1479 \]
 different 
random samples. High-precision measurements of various critical properties
yielded
$T_c = 3.49921(3), \alpha/\nu=-0.066(9), \beta/\nu=0.505(2), 
\gamma/\nu=1.990(4), 1/\nu = 1.467(5)$, in good agreement with\cite{heuer}.

{\it Lack of self-averaging, distribution of the critical susceptibility:}  
Fig 1. presents curves of $\chi(T)$
vs temperature for two
samples of size $L=32$. 
\begin{figure}[h]
      \epsfxsize=73mm
      \centerline{ \epsffile{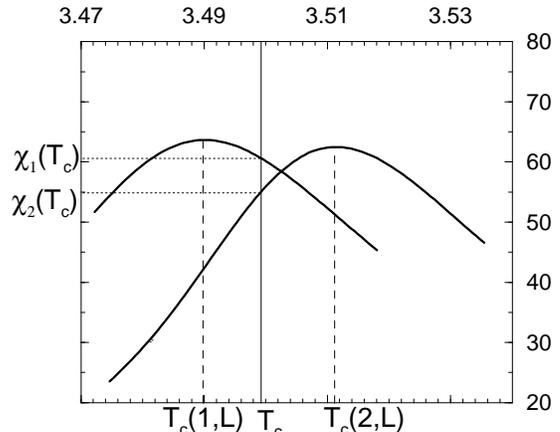} }
      \caption{ 
Susceptibility per spin of two systems of size $L=32$, generated with 
site occupation probability $p=0.8$, versus temperature. 
$T_c(i,L)$ denotes the pseudocritical temperature of sample $i$,
identified by the maximum of the susceptibility. $T_c$ is the 
value approached by the mean of these in the $L \rightarrow \infty$
limit. $\chi_i(T_c)$ are the values of $\chi$ as measured at $T_c$
for sample $i$.
  }
\label{fig: curves}
\end{figure}
The values of $\chi_i(T_c)$ and the pseudocritical
temperatures of the two samples are clearly identified. From measuring
$\chi_i(T_c)$ for an ensemble of samples we constructed the histogram
of Fig 2, for various sizes $L$.
\begin{figure}[h]
      \epsfxsize=85mm
      \centerline{ \epsffile{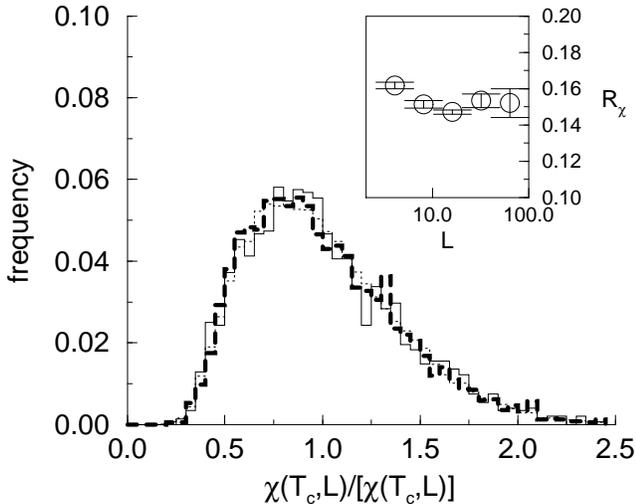} }
      \caption{
Distribution of the critical susceptibilities, measured at
$T_c$ for different samples and different sizes $L=16$ (thin dotted line),
$L=32$ (thick dashed) and $L=64$ (thin solid).  For each
$L$ the values of $\chi_i(T_c,L)$ were normalized by the ensemble average
of all samples of that size. The inset shows the second moment of the
distribution as a function of $L$; evidently $R_\chi \rightarrow C$.
 }
\label{fig: Xic.prob}
\end{figure}
 As evident from the histograms and
from the inset, the widths of these distibutions approach
a constant for increasing $L$. This is the result of AH,
whereas our scaling theory\cite{wd95} would predict
$R_{\chi} \sim L^{\alpha/\nu}$. The source of the discrepancy is an 
assumption
we made regarding the distribution of the pseudocritical temperatures; had we
assumed that (\ref{eq:AHwidth}) holds, our scaling ansatz would have also
produced the correct result $R_\chi \sim$const. Therefore confirmation 
of this
result may serve as an indication of the validity of (\ref{eq:AHwidth}) 
as well.
Nevertheless, since this interpretation depends on the validity of 
further assumptions implicit in our scaling theory, it is important to 
verify independently and directly scaling and (\ref{eq:AHwidth}).

{\it Universality:} We studied two more random ensembles: another grand 
canonical
one, but with occupation probability $p=0.6$ and a canonical one 
with a fixed 
number of  $0.6 L^d$ spins, randomly placed in each sample.     We found that 
the limiting values of both $R_\chi$ and $R_m$ are the same for the two
grand-canonical ensembles (with $p=0.8, 0.6$) indicating universality, but
for the canonical ensemble different values were obtained. These results
will be presented and explained elsewhere\cite{ahw98}.

{\it Distribution of the pseudocritical temperatures:}
We estimated, using the histogram reweighting method\cite{Swendsen 93},
the pseudocritical temperature $T_c(i,L)$ of
every sample of our ensembles. This was done in an iterative way; the first
guess for $T_c(i,L)$ was $T_c$ - from  data collected at this point we 
calculated the susceptibility
as a function of $T$; the temperature at which it had its maximum 
was our next estimate for $T_c(i,L)$ where more data were collected and 
so on.
The procedure converged in less than ten iterations within the resolution
imposed by the statistical error in determining the temperature of the 
maximal
$\chi$. The resulting histograms for several sizes are collapsed on Fig 3
using the value of $y_t=\frac{1}{\nu}=1.467(5)$ which was determined independently\cite{wd98} by simulations at $T_c$.
\begin{figure}
      \epsfxsize=85mm
      \centerline{ \epsffile{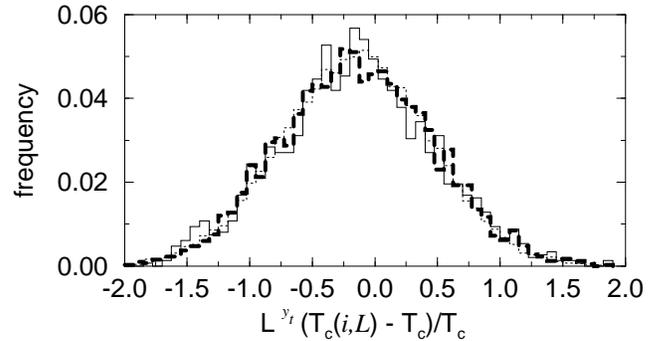} }
      \caption{ 
Distribution of scaled pseudocritical temperatures obtained for
sizes $L=16$ (thin dotted 
 line), $L=32$ (thick dashed) and $L=64$ (thin solid), using 
$y_t=1/\nu=1.467$ and $T_c=3.4992$.
}
\label{fig: Tc.prob}
\end{figure}
The means and widths of these distributions are 
\begin{eqnarray}
T_c(L) & \approx &T_c -  0.532 \cdot L^{-{1/\nu}} \nonumber \\  
\delta T_c(L) & \approx & 2.13 \cdot L^{-{1/\nu}}
\label{Tdist}
\end{eqnarray}
The width of the distribution exceeds by a factor of nearly 4 the shift
of its average  from $T_c$. Hence the large fluctuations in $\chi(T_c)$
are due to the large value of $ \delta T_c(L) $; when we perform 
measurements
at the fixed temperature $T_c$, some samples we deal with will be 
considerably  above  their pseudocritical temperature and some below!
(see Fig 1) 
 We have also made 
straightforward fits of
the variance $ \delta T_c(L)^2 $ to the form 
$ \delta T_c(L)^2 \sim L^{-2\rho}$, and fits of the shift of $T_c(L)$ to
$T_c(L)-T_c \sim l^{-\lambda}$; we found $\rho=1.449(8)$ and $\lambda=1.42(4)$.
Clearly $\rho$ is within errors of the shift exponent $\lambda$ and its error
 bar is 
small enough to conclude that $\rho\not= d/2=1.5$. Since one expects that
$\lambda=y_t$ and we have also determined independently $y_t=1.467(5)$
our results strongly suggest that all these exponents are actualy the same 
$\rho=\lambda=1/\nu$. 

{\it Testing the scaling ansatz:} Eq. (\ref{eq:FSS}) represents an implicit 
assumption; that the most important effect of the sample dependence of each
realization of the randomness can be absorbed in the pseudocritical
temperature of the particular sample. Our first task was to check the extent
to which this holds, i.e. to what extent can one collapse data obtained
at different
sizes and temperatures for different samples? Fig. 4 presents the 
magnetization
for two system sizes.
 The data collapse quite well onto two branches of
a function, one below and one above $T_c(i,L)$, lending support to the
validity of the scaling ansatz. We succeeded to fit thousands of data points at
three different sizes to two scaling functions (corresponding to the two
branches). This may be interpreted as supporting a {\it strong scaling
hypothesis}, e.g. that for large $L$
\beq
X_i(T,L) \approx L^\rho {\tilde Q}({\dot t_i}L^\frac{1}{\nu})   
\label{eq:strong}
\eeq
Here we modified (\ref{eq:FSS}) by dropping the sample-dependence of
the scaling function, i.e. assuming that the {\it entire} sample dependence
can be absorbed in ${\dot t_i}$. On the other hand, as evident from Fig. 4,
the data points  do exhibit considerable scatter about the main trend.
To investigate the extent of violation of such a strong scaling relation,
we studied the distribution of the susceptibility maxima. 
\begin{figure}[h]
      \epsfxsize=70mm
      \centerline{ \epsffile{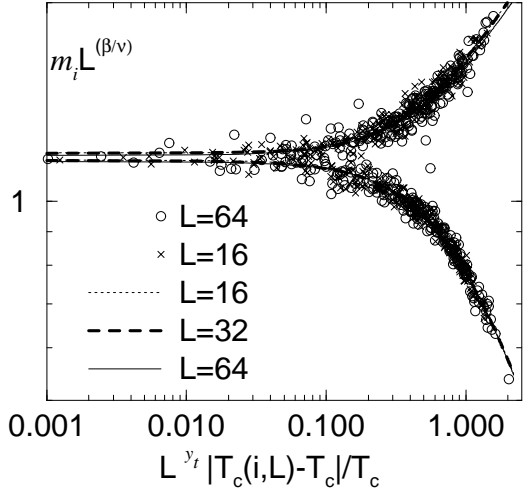} }
      \caption{
Scatter plot of the scaled magnetization, measured 
at $T_c$ for ensembles of samples of two
different sizes, plotted versus the scaled temperature
${\dot t}_i=\{T_c-T_c(i,L)\}/T_c$. The lines are
the scaling functions fitted separately for three different sizes,
showing remarkable agreement. The scatter of the points from
the lines is due partially to measurement error (thermal) and partly 
to genuine sample dependence of the scaling function.
  }
\label{fig: mdata.fit}
\end{figure}

{\it Sample dependence of the scaling function:}
If, instead of $\chi_i(T_c,L)$,
we  measure for each sample and size the {\it maximal} value of the
susceptibility, $\chi^{max}_i=\chi[T_c(i,L),L)]$, we accumulate all our data
at the same value of the scaling variable $ {\dot t_i}L^\frac{1}{\nu}=0$.
Hence if for large $L$ the scaling function 
goes to a sample independent form, the distribution of $\chi^{max}_i$ should
approach a $\delta$ function. This distribution is presented in Fig 5:
even though it is much narrower (by a factor of $\approx 70$)
than the distribution of Fig 2, as the inset
shows, its width also {\it goes to a constant}. Hence a strong scaling
hypothesis such as (\ref{eq:strong}) can be viewed only as an approximation.

{\it Consequences for efficient simulations}: In order to acquire data for
large systems with high precision (say to estimate exponents from 
finite size scaling analysis) the commonly accepted procedure is to perform,
for all sizes, simulations at one temperature, $T_c$. The error inherent in
doing this, due to sample to sample fluctuations, is much larger than that
of taking measurements at $T_c(i,L)$. The latter procedure involves a 
different computational overhead, that of determining   $T_c(i,L)$. Clearly,
this may be advantageous if the width of the quantities measured at the
pseudocritical temperature of each sample is much smaller than that of
the data taken at $T_c$.
\begin{figure}[h]
      \epsfxsize=85mm
      \centerline{ \epsffile{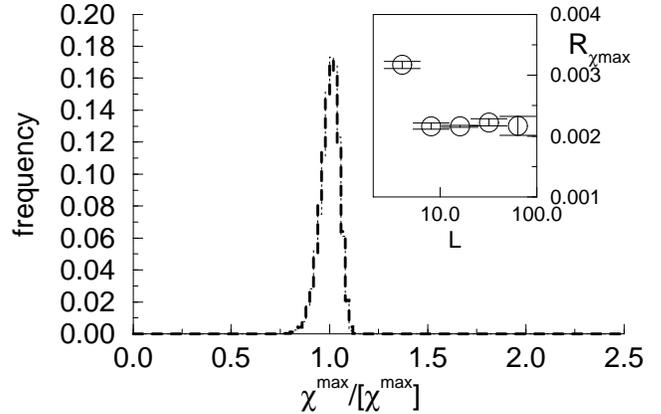} }
      \caption{
Distribution of the normalized maximal susceptibilities
for samples of sizes $L=16$ (thin dotted line) and $L=32$ (thick dashed).
 The horizontal scale is the same as Fig 2,
to emphasize how small is the width of this distribution, which also
goes to a constant (see inset).
}
\label{fig: Xmax.prob2}
\end{figure}

We thank A. Aharony,
 A.B. Harris, W. Janke, M. Picco and
 D. Stauffer
  for useful discussions and suggestions,
H.O. Heuer for correspondence and D. Lellouch for advice on statistics.  
This research has been supported in part by the Germany-Israel Science
 Foundation (GIF) and in part by the Israel Ministry of Science.
Computations were performed  on the SP2 at the Inter-University High
 Performance Computing Center, Tel Aviv.

\end{document}